\documentstyle[12pt]{article}
\begin{document}
\date { }

\centerline { \bf { Division of Differential operators , intertwine relations}}
\centerline {
\bf {  and Darboux Transformations  }} \centerline { \bf
{ }}
\centerline {  {A.A. Zaitsev,S.B. Leble}}
Theoretical Physics Department, Kaliningrad State University, 236041 Kaliningrad,

Al.Nevsky st. 14.

Theoretical l Physics and Mathematical Methods Department, 
Technical University 
of Gdansk, ul. Narutowicza, 11/12. 80-952, Gdansk-Wrzeszca, Polamd. 

email leble@mifgate.pg.gda.pl
\medskip
\abstract
The problem of a differential operator left- and right division is solved in terms of generalized
 Bell polinomials for nonabelian differential unitary ring . The definition of the polinomials is
made by means of  recurrent relations. The expresions of classic Bell polinomils via generalized
one is given. The conditions of an exact factorization possibility leads to the intertwine
relation and results in some linearizablegeneralized Burgers equation. An alternative
proof of the Matveev theorem is given and  Darboux - Matveev transformations formula for
coefficients follows from the intertwine relations and also expressed in the generalized
Bell polinomials. 

\bigskip
\section
{ \bf  Introduction}

\medskip
Problems of solitonics and, more general, of nonlinear partial differential equation
(NLPDE) integration is simplifying when one introduce
 a special factorization of a (differential) linear operator associated with the problem. The
technique was actively used in theoretical physics [1] as well as in mathematics from early [2] to
[3]. First comes from attempts to build explicit solutions of the one-dimensional Schr\"odinger
equation and leads to so-called "supersymmetry" and the second relates to general problems of a
differential field theory. The "physical line" was developed for partial differential
equations (PDE) in [4]  where the direct connection with the theory of Darboux
transformations was noticed. Both approaches connect factors with some Riccati
equation that is a feature of the general factorization problem [5] of importance from
algorithmic point of view [6]. The problem of factorization was succesfully investigated
algebraically for a commutative field in [7,8] and nonabelian one in [9] that allows to
find appropriate algorithm for computations [10].  Applications to nonlinear ordinary
differential equations (NODE) decomposition theory shows a way to NLPDE via classical
reductions of S. Lie (point symmetries).
	A search of approaches to solution of integrable NLDPE or  looking for somewhat sort of a
 language in which algebraic algorithms could be formulated in a compact form lead us to
the theory of Bell polinomials. The polinomials appears naturally in Darboux techniques
[11] and even a disarmed eye see a special polinomial combinations of derivatives in famous
integrable equations. This way having in mind factorization of Schr\"odinger operators and
Cole-Hopf linearization of the Burgers equation the Bell polinomials in fact appear in
many contexts and generalized for nonabelian coefficients at [12]. The last result leads
to a nonabelian Burgers hierarchy, directly linearized by the Cole-Hopf substitution
analogue.
	From the side of integrable nonlinear differential equations form and their bilinear (Hirota)
 counterparts the introduction of "binary" Bell polinomials allows to express the algorithm of
linearization of a given NLDPE in compact combinatoric language [13].

In the Section 2 of this paper we, after some notations, introduce left and right
nonabelian polinomials by recurrence relations. The polimomials correspond to [12] but
the form of their introduction is adjusted to our purposes and give some new useful
relations. We also prove similar relationships in our context. 
	In the Secton 3 we make next step on a way of the generalization of bell polinomials. We
introduce some auxiliary DO and show how to make a linear differential operator (LDO)
division by left and right factors of the first order LDO and  formulate conditions in
which a   factorization is possible (give a zero remainder) for a nonabelian field. Next
section (4) links a solution of the problem of a factorization with generalized (and
linearized by Cole-Hopf substitution analog) Riccati equation. At the last section we give
a compact formulation and proof of the Matveev theorem [15,16] in our notations deriving
the potenyials' (LDO coefficients) transforms in term sof Bell polinomials.

\section 
{\bf Main notations and auxiliary results. Bell polinomials.}

 Let K be a differential ring of the zero characteristics with unit e (i.e. unitary ring) and an 
involution denoted by a superscript *. The differentiation is denoted as D. The differentiation and
the involution are in accordance with operations in K, i.e.

1.$(a^*)^* = a, (a + b)^* = a^* + b^*; (ab)^* = b^*a^*, a,b \in K$

2.$D(a+b) = Da + Db; D(ab) = (Da)b + aDb;$

$$3. (Da)^*  = - Da^*; \eqno (1)$$

4. the $D^n$ operators form a basis in K-module Diff(K) of differential operators.

The subring of constants is $ K_0$ and a multiplicative group of elements is G.

5. For any $s\in K$ there exists an element $\varphi\in K$ such that $D\varphi = s\varphi$,
that means also the existence of a solution of the equation 
$$D\phi = - \phi s  \eqno (2)$$ 
due to the involution properties.

There are lot of applications in the theory of integrable nonlinear equations and
 quantum problems connected with  rings of square matrices. In this case the matrices are
parametrized  by a variable $x$ and D may be a derivative with respect to this variable or
a combination of partial derivatives that satisfy the conditions (3,4). If D is such usual
differentiation, then the involution "*" is the Hermitian conjugation.In the case of a
commutator $Da = [d,a]$, $(Da)^* = - [d^*,a]$.  Having in mind such or similar applications
we shall name the involution as conjugation. We do not restrict oneself by the
matrix-valued case: an appropriate operator ring also good for our theory.

Below we introduce left and right nonabelian  Bell poplinomials (see also [12]
The left differential Bell polinomials are
defined by
\medskip
{Definiton 1}
$$B_0(s) = e \eqno(3)$$
and the recurrent relation
$$B_n(s) = DB_{n-1}(s) + B_{n-1}(s)s, n =
1,2,..\eqno(4)$$

{\bf{Statement 1}}.\enspace {\it If the element}\enspace
$\varphi\,\in\,G$\enspace {\it{satisfy the equation}}\, $D\varphi = s\varphi$
{\it{then}}
$$D^n\,\varphi=B_n(s)\,\varphi,\quad n=0,1,2,\dots.
\eqno(5)$$
\medskip
{\bf{Proof}}.\enspace Let us use the induction. At $n=0,1$
\enspace the equality (5) is trivial, from (3,4)\enspace $n=1$\enspace
 we get
\medskip

$$B_1(s)=s;$$
 \medskip
therefore,
$$D\,\varphi=s\,\varphi=B_1(s)\,\varphi.
\eqno(6)$$
\medskip
Here and further we shall denote $Dx = x', x \in K$ for brevity especially when the
operator D acts only on the nearest element.
\medskip
 Let the equality (5) be trivial for some \, $n$,\, then, by
means of the relations (6), (8) we obtain
\medskip

$$D^{n+1}\,\varphi=\,D(B_n(s)\,\varphi)=
B_n'(s)\,\varphi+B_n(s)\,D\,\varphi=
B_n'(s)\,\varphi+B_n(s)\,s\,\varphi=$$
$$(D\,B_n(s)+B_n(s)\,s)\,\varphi=B_{n+1}(s)\,\varphi;$$
it means that (5) take place for the next value of \, $n+1$, hence for all the natural \,
$n$ or zero.\quad
$\Box$
\medskip

Evaluations by (6) give
\medskip

$$B_1(s)=s;\enspace B_2(s)=s^2+Ds;\enspace
B_3(s)=s^3+2s'\,s+s\,Ds+D^2s;$$
$$B_4(s)=s^4+3s'\,s^2+2s\,s'\,s+s^2\,Ds+3s''\,s+3(Ds)^2+s\,D^2s+D^3s.$$
\medskip

Right Bell polinomials are introduced similarily. 
\medskip

{\bf Definition 2 }{\it{The right nonabelian differential Bell polinomials }}\enspace
are defined by the equalty
\medskip

$$B_0^+(s)=e
\eqno(7)$$
and \enspace {\it{a recurrence}}

$$B_n^+(s)=-DB_{n-1}^+(s)+s\,B_{n-1}^+(s),\quad n=1,2,\dots.
\eqno(8)$$
\medskip

{\bf{Statement 2}}.\enspace {\it{If the element}}\enspace
$\phi\,\in\,G$\enspace {\it{satisfies the equation}}\, (2),\,
{\it{then}}

$$D^n\,\phi=(-1)^n\,\phi\,B_n^+(s),\quad n=0,1,2,\dots.
\eqno(9)$$
\medskip

{\bf{Proof}}.\enspace Treat the induction again. At\enspace
$n=0,\,1$\enspace the equality (8) trivializes and \enspace $n=1$
\enspace yields (2), whence (7),
(8) give
\medskip

$$B_1^+(s)=s;$$
Otherwise,
\medskip

$$D\,\phi=-\phi\,s=(-1)^1\,\phi\,B_1^+(s).
 $$
\medskip

Let the equality  (9) is valid for some\, $n$,\, then, by means of (2), (8), (10) we get
\medskip

$$D^{n+1}\,\phi=D((-1)^n\,\phi\,B_n^+(s))=
(-1)^n\,(\phi'\,B_n^+(s)+\phi\,DB_n^+(s))=$$
$$(-1)^n\,(-\phi\,s\,B_n^+(s)+\phi\,DB_n^+(s))=
(-1)^{n+1}\,\phi\,(-DB_n^+(s)+s\,B_n^+(s))=$$
$$(-1)^{n+1}\,\phi\,B_{n+1}^+(s);$$
so, it is valid  for the next\, $n+1$,\, hence for any
natural or zero $n$.\quad $\Box$
\medskip

Calculations by (8) give
\medskip

$$B_1^+(s)=s;\enspace B_2^+(s)=s^2-Ds;\enspace
B_3^+(s)=s^3-s'\,s-2s\,Ds+D^2s;$$
$$B_4^+(s)=s^4-s'\,s^2-2s\,s'\,s-3s^2\,Ds+s''\,s+3(Ds)^2+3s\,D^2s-D^3s.
$$
\medskip
The right polinomials go to ones determined at [12] if change $D \rightarrow - D$ .
There exists a simple but useful link between the right and left Bell polinomials generated by
the {\it{conjugation}}.
\medskip

{\bf{Statement 3}}.\enspace {\it{The left and right Bell polinomials are connected by the
following relations}}:
\medskip

$$B_n(s)^*=B_n^+(s^*),\quad B_n^+(s)^*=B_n(s^*).
\eqno(11)$$
\medskip

{\bf{Proof}}.\enspace It is enough to prove the first one as the second is conjugate. 
For this
proof the induction again 
 starts from a trivial point
\enspace $n=0$\enspace of the equality (11 ), use the recurrences (4), (8) and give
\medskip

$$B_{n+1}(s)^*= -DB_n(s)^*+s^*\,B_n(s)^*=
-DB_n^+(s^*)+s^*\,B_n^+(s^*)= B_{n+1}^+(s^*).$$
That means the validity for any\, $n$.\quad $\Box$
\medskip

If the ring is abelian, left and right polinomials coincide..

\medskip

{\bf{Remark \, 1}}.\enspace {\bf{The statement 4}}\enspace means,
that for the Bell polinomials takes place \enspace {\it{a duality}}:
\enspace {\it{any relation connecting right BP  goes to one connecting the left BP and vice
versa}}. Let us denote further
\medskip

$$L_s=D-s.
\eqno(12)$$
\medskip

Note that the recursion (7)  may be written by means
of the designation (12) as
\medskip

$$B_{n+1}^+(s)=-L_s\,B_n^+(s),\quad n=0,1,2,\dots;
 $$
with the simple corollary
\medskip
$$B_n^+(s)=(-1)^n\,L_s^n\,e,\quad n=0,1,2,\dots.
\eqno(13)$$

\medskip
\section{ One more generalization of Bell polinimials}

In the next section the problem of division of an arbitrary operator \, $L$\, by the operator\,
$L_s$ will be studied.\enspace For the solution of it,
we introduce auxiliary operators for a right division\, $H_n$\, by means of
\medskip

{\bf Definition 3}
$$H_0=e.
\eqno(14)$$
and\enspace {\it{the recurrence:}}:
\medskip

$$H_n=D\,H_{n-1}+B_n(s),\quad n=1,\,2,\,\dots.
\eqno(15)$$
\medskip

{\bf{Statement 5}}.\enspace {\it{The identity holds}}
\medskip

$$D^n=H_{n-1}\,L_s+B_n(s),\quad n=1,2,\dots.
\eqno(16)$$
\medskip

{\bf{Proof}}.\enspace The induction gives the following. At\enspace $n=1$
\enspace the equality (16) is trivial. Let it be for \, $n$,\, then

$$D^{n+1}=D\,(H_{n-1}\,L_s+B_n(s))=D\,H_{n-1}\,L_s+D\,B_n(s)=
D\,H_{n-1}\,L_s+B_n(s)\,D+DB_n(s)=$$
$$D\,H_{n-1}\,L_s+B_n(s)\,L_s+DB_n(s)+B_n(s)\,s=
(D\,H_{n-1}+B_n(s))\,L_s+B_{n+1}(s)=$$
$$H_n\,L_s+B_{n+1}(s),$$
it is (16), but after the change n for $n+1$.
\enspace Hence, the equality (16) is valid for all natural\, $n$.\enspace $\Box$
\medskip

Coefficients of the operators \, $H_n$\,are expressed via\enspace
{\it{generalized Bell polinomials }},\enspace that are defined by
\medskip

{bf Definition 4}
$$B_{n,\,0}(s)=e,\quad n=0,1,2,\dots,
\eqno(17)$$
and recurrence relations
\medskip

$$B_{n,\,k}(s)=B_{n-1,\,k}(s)+DB_{n-1,\,k-1}(s),\quad
k=\overline{1,\,n-1},\quad n=2,3,\dots.
\eqno(18)$$
\medskip

$$B_{n,\,n}(s)=DB_{n-1,\,n-1}(s)+B_n(s),\quad n=1,2,\dots.
\eqno(19)$$
\medskip

{\bf{Statement\, 6}}.\enspace {\it{Generalized Bell polinomials are coefficients of the
operators}}\,
$H_n$,\, {\it{i.e.}}
\medskip

$$H_n=\sum_{k=0}^n\,B_{n,\,n-k}(s)\,D^k,\quad n=0,1,2,\dots.
\eqno(20)$$
\medskip

{\bf{Proof}}.\enspace When\enspace $n=0$\enspace the equality (22)
trivializes\enspace $H_0=e$.\enspace
so that it is enough to establish for operators
(20) the recurrence (15), by the following transformations:
\medskip

$$H_n-D\,H_{n-1}-B_n(s)=\sum_{k=0}^n\,B_{n,\,n-k}(s)\,D^k-
\sum_{k=0}^{n-1}\,D\,B_{n-1,\,n-1-k}(s)\,D^k-B_n(s)=$$
$$\sum_{k=0}^n\,B_{n,\,n-k}(s)\,D^k-\sum_{k=0}^{n-1}\,
(B_{n-1,\,n-1-k}(s)\,D^{k+1}+DB_{n-1,\,n-1-k}(s)\,D^k)-B_n(s)=$$
$$\sum_{k=0}^n\,B_{n,\,n-k}(s)\,D^k-\sum_{k=0}^{n-1}\,
B_{n-1,\,n-1-k}(s)\,D^{k+1}-\sum_{k=0}^{n-1}\,DB_{n-1,\,n-1-k}(s)\,D^k-
B_n(s)=$$
$$\sum_{k=0}^n\,B_{n,\,n-k}(s)\,D^k-\sum_{k=1}^n\,B_{n-1,\,n-k}(s)\,D^k-
\sum_{k=0}^{n-1}\,DB_{n-1,\,n-1-k}(s)\,D^k-B_n(s)=$$
$$(B_{n,\,0}(s)-B_{n-1,\,0})\,D^n+\sum_{k=1}^{n-1}\,(B_{n,\,n-k}(s)-
B_{n-1,\,n-k}(s)-DB_{n-1,\,n-1-k}(s))\,D^k+$$
$$(B_{n,\,n}(s)-DB_{n-1,\,n-1}(s)-B_n(s))=(e-e)\,D^n=0.$$
\medskip

As the equality (14) and the recurrence (15) define the operators
 \, $H_n$,\, in unique way, the result of the calculations means the validity of
(20).\enspace
$\Box$
\medskip

The formulas  (18) and  (19) are simple, but nor very convenient for evaluations of
\enspace $B_{n,\,k}(s)$
\newline
(you forced to go "ladder" way),therefore we suggest more complicated but practically
easier algorithm. For this we put the representation (20)
into (18); getting 
\medskip

$$0=D^{n+1}-H_n\,L_s-B_{n+1}(s)=
D^{n+1}-\sum_{k=0}^n\,B_{n,\,n-k}(s)\,D^k\,(D-s)-B_{n+1}(s)=$$
$$D^{n+1}-\sum_{k=0}^n\,B_{n,\,n-k}(s)\,D^{k+1}+
\sum_{k=0}^n\,B_{n,\,n-k}(s)\,D^k\,s-B_{n+1}(s)=$$
$$D^{n+1}-\sum_{k=1}^{n+1}\,B_{n,\,n-k+1}(s)\,D^k+\sum_{k=0}^n\,
B_{n,\,n-k}(s)\,\sum_{i=0}^k\,{k \choose i}\,D^{k-i}s\,D^i-B_{n+1}(s)=
$$
$$(e-B_{n,\,0}(s))\,D^{n+1}-\sum_{k=1}^n\,B_{n,\,n-k+1}(s)\,D^k+
\sum_{0\,\le\,i\,\le\,k\,\le\,n}\,{k \choose i}\,B_{n,\,n-k}(s)\,
D^{k-i}s\,D^i-B_{n+1}(s)=$$
$$(e-B_{n,\,0}(s))\,D^{n+1}-\sum_{k=1}^n\,B_{n,\,n-k+1}(s)\,D^k+
\sum_{0\,\le\,k\,\le\,i\,\le\,n}\,{i \choose k}\,B_{n,\,n-i}(s)\,
D^{i-k}s\,D^k-B_{n+1}(s)=$$
$$(e-e)\,D^{n+1}-\sum_{k=1}^n\,B_{n,\,n-k+1}(s)\,D^k+\sum_{k=0}^n\,
\sum_{i=k}^n\,{i \choose k}\,B_{n,\,n-i}(s)\,D^{i-k}s\,D^k-B_{n+1}(s)=
$$
$$\sum_{k=1}^n\,\left(-B_{n,\,n-k+1}(s)+\sum_{i=k}^n\,{i \choose k}\,
B_{n,\,n-i}(s)\,D^{i-k}s\right)\,D^k+\left(\sum_{i=0}^n\,
B_{n,\,n-i}(s)\,D^is-B_{n+1}(s)\right).$$
\medskip

The following  formulas are extracted:
\medskip

$$B_{n,\,n-k+1}(s)=\sum_{i=k}^n\,{i \choose k}\,B_{n,\,n-i}(s)\,
D^{i-k}s,\quad k=\overline{1,\,n},\enspace n=0,1,2,\dots;
\eqno(21)$$

$$B_{n+1}(s)=\sum_{i=0}^n\,B_{n,\,n-i}(s)\,D^is,\quad n=0,1,2,\dots.
\eqno(22)$$
\medskip

The formulae (22) expresses the standard (nonabelian) Bell polinomials via the generalized
ones; it may be rewrited as 
\medskip

$$B_{n+1}(s)=\sum_{i=0}^n\,B_{n,\,i}(s)\,D^{n-i}s,\quad n=0,1,2,\dots.
\eqno(22)$$
\medskip

If in the formulae (22) rearrange the summation by\enspace
$k\,\rightarrow\,n-k+1$,\enspace then after simple work yields:
\medskip

$$B_{n,\,k}(s)=\sum_{i=0}^{k-1}\,{n-i \choose n-k+1}\,B_{n,\,i}(s)\,
D^{k-i-1}s,\quad k=\overline{1,\,n},\enspace n=0,1,2,\dots.
\eqno(23)$$
\medskip

Evaluation of the generalized Bell polinomials by (22) gives
\medskip

$$B_{n,\,1}(s)=s;\quad B_{n,\,2}(s)=s^2+n\,Ds;\quad
B_{n,\,3}(s)=s^3+n\,s'\,s+(n-1)\,s\,Ds+{n \choose 2}\,D^2s;$$
$$B_{n,\,4}(s)=s^4+n\,s'\,s^2+(n-1)\,s\,s'\,s+(n-2)\,s^2\,Ds+
{n \choose 2}\,s''\,s+n(n-2)\,(Ds)^2+$$
$${n-1 \choose 2}\,s\,D^2s+{n \choose 3}\,D^3s.$$
\medskip

For a solution of problem of the left division of a differential operator, $L$\,
by \, $L_s$\,  the similar but simpler consideration is necessary. The analog of \enspace
the {\bf{Statement 5}}\enspace is
\medskip

{\bf{Statement 7}}.\enspace {\it{The following identity is valid}}
\medskip

$$D^n=L_s\,H_{n-1}^++B_n^+(s),\quad n=1,2,\dots,
\eqno(27)$$
{\it{where}}
\medskip

$$H_n^+=\sum_{k=0}^n\,B_{n-k}^+(s)\,D^k,\quad
n=0,1,2,\dots.
\eqno(28)$$
\medskip

{\bf{Proof}}.\enspace Let us solve the problem of the left division of the operator\, $D^n$\,
by the operator \, $L_s$\,i.e. put
\medskip

$$D^n=L_s\,H_{n-1}^++r,
\eqno(29)$$
where \, $r$ --remainder, and
\medskip

$$H_{n-1}^+=\sum_{k=0}^{n-1}\,b_k\,D^k.
\eqno(30)$$
The remainder, $r$\, and the coefficients $\,b_k\,$ are to be determined.

Substituting (30) into (29)
and, transforming, we arrive at
\medskip

$$L_s\,H_{n-1}^-+r=(D-s)\,\sum_{k=0}^{n-1}\,b_k\,D^k+r=
\sum_{k=0}^{n-1}\,D\,b_k\,D^k-\sum_{k=0}^{n-1}\,s\,b_k\,D^k+r=$$
$$\sum_{k=0}^{n-1}\,(b_k\,D^{k+1}+Db_k\,D^k)-
\sum_{k=0}^{n-1}\,s\,b_k\,D^k+r=
\sum_{k=0}^{n-1}\,b_k\,D^{k+1}+\sum_{k=0}^{n-1}\,Db_k\,D^k-
\sum_{k=0}^{n-1}\,s\,b_k\,D^k+r=$$
$$\sum_{k=1}^n\,b_{k-1}\,D^k+\sum_{k=0}^{n-1}\,(Db_k-s\,b_k)\,D^k+r=
(Db_0-s\,b_0+r)+\sum_{k=1}^{n-1}\,(b_{k-1}+Db_k-s\,b_k)\,D^k+
b_{n-1}\,D^n=$$
$$(L_s\,b_0+r)+\sum_{k=1}^{n-1}\,(b_{k-1}+L_s\,b_k)\,D^k+b_{n-1}\,D^n.$$
\medskip
Comparing the result with the left-hand side of (31),we obtain
\medskip

$$b_{n-1}=e;\quad b_{k-1}=-L_s\,b_k,\quad k=\overline{1,\,n-1};\quad
r=-L_s\,b_0.$$
Recalling (7) and the recurrence (8), we represents:
\medskip

$$b_k=B_{n-k-1}^+(s),\quad k=\overline{0,\,n-1};\quad r=B_n^+(s).$$
\medskip
The equalities  (24) and (25) are their obvious corollaries.\quad $\Box$
\bigskip

{\bf{3.\enspace Division and factorization of differential operators.
\newline
Generalized Riccati equations}}.
\medskip

Let  
\medskip

$$L=\sum_{n=0}^N\,a_n\,D^n,\quad a_n\,\in\,K,\, -
\eqno(28)$$
be a differential operator of the order\, $N$.\, We shall study a right and left division of\,
$L$\, by the operator \, $L_s, $,\,defined by the equality (12), Suppose
\medskip

$$L=M\,L_s+r,
\eqno(29)$$

$$L=L_s\,M^++r^+,
\eqno(30)$$
where \enspace $M,\, M^+$ --results of right and left division and \enspace
$r,\, r^+$ -- remainders.
\medskip

The statement 5 allows to solve the problem in a simple way.
\medskip

{\bf{Statement 8}}.\enspace {\it{If the representation}}\,
(29),\enspace {\it{is valid, then for the remainder }}\, $r$\, {\it{and the result}}\, $M$
\enspace {\it{yields}}:
\medskip

$$r=\sum_{n=0}^N\,a_n\,B_n(s);
\eqno(31)$$

$$M=\sum_{n=1}^N\,a_n\,H_{n-1},
\eqno(32)$$
{\it{or}}
\medskip

$$M=\sum_{n=0}^{N-1}\,b_n\,D^n,
\eqno(33)$$
{\it{where}}

$$b_n=\sum_{k=n+1}^N\,a_k\,B_{k-1,\,n}(s),\quad n=\overline{0,\,N-1}.
\eqno(34)$$
\medskip

{\bf{Proof}}.\enspace Multiplying both sides of the equality (16)
by\, $a_n$\, and summing by\, $n$,\,one get (29), (31) -- (33).
After the substitution of the representation (22) into (34) it results
\medskip

$$M=\sum_{n=1}^N\,a_n\,H_{n-1}=\sum_{n=1}^N\,a_n\,\sum_{k=0}^{n-1}\,
B_{n-1,\,k}(s)\,D^k=
\sum_{1\,\le\,n\,\le\,N,\,0\,\le\,k\,\le\,n-1}\,a_n\,B_{n-1,\,k}(s)\,
D^k=$$
$$\sum_{1\,\le\,k\,\le\,N,\,0\,\le\,n\,\le\,k-1}\,a_k\,B_{k-1,\,n}(s)\,
D^n=\sum_{n=0}^{N-1}\,\sum_{k=n+1}^N\,a_k\,B_{k-1,\,n}(s)\,D^n,$$
from what the formulas (33), (34) follow.
\medskip

As a corollary one get
\medskip

{\bf{Statement 9}}.\enspace {\it{For the linear operator}}\, $L$\, {\it{to be a right divisible
by}}\,
$L_s$\, {\it{without remainder}}\, it is\, {\it{necessary and enough,
that}}\, $s$\, {\it{to be a solution of the differential equation}}\enspace
 
\medskip

$$\sum_{n=0}^N\,a_n\,B_n(s)=0.
\eqno(35)$$

{\it{If this condition holds, the operator}}\, $L$\,
{\it{factors as }}:
\medskip

$$L=M\,L_s,$$
{\it{where the value of the operator }}\, $M$\, {\it{is given by}}\, (32)\,
{\it{or by the expressions}}\, (33), (34).
\medskip

The equation (35) is nonlinear. At\, $N=2$\, it is Riccati equation,
therefore it is natural to name it as a  \enspace
{\it{generalized right Riccati equation}}.
The right Riccati equation is generalized by means of the Statement 2.
\medskip

{\bf{Statement 10}}.\enspace {\it{Let an invertible function}}\,
$\varphi$\, {\it{be a solution to the linear differential equation}}
\medskip

$$\sum_{n=0}^N\,a_n\,D^n\,\varphi=0.
\eqno(36)$$
{\it{Then the operator}}\, $L$,\, {\it{defined by the equality}}\, (31),\,
{\it{is right divisible by}}\, $L_s$,\, {\it{where}}
\medskip

$$s=\varphi'\,\varphi^{-1}.$$
\medskip

{\it{Morover}}\, $s$\, {\it{is a solution of the right Riccati equation}}\, (35).

For the solution of a left-division problem letus write a result in a form
\medskip

$$M^+=\sum_{n=0}^{N-1}\,b_n^+\,D^n,
\eqno(37)$$
looking for the determination of \enspace $b_n^+,\enspace
n=\overline{0,\,n-1}$.

Substitute the representation (37) into the right-hand side of 
the equation (30). By means the calculus analogues to that was used for the proof of the
Statement 7, you obtain:
\medskip

$$b_{N-1}^+=a_N;
\eqno(38)$$

$$b_n^+=a_{n+1}-L_s\,b_{n+1}^+,\quad n=\overline{0,\,N-2};
\eqno(39)$$

$$r^+=a_0-L_s\,b_0^-.
\eqno(40)$$
\medskip

Solving subsequently the equations
(37) -- (40) one goes to 
\medskip

$$b_n^+=\sum_{k=n+1}^N\,(-1)^{k-n-1}\,L_s^{k-n-1}\,a_k,\quad
n=\overline{0,\,N-1};
\eqno(41)$$

$$r^+=\sum_{k=0}^N\,(-1)^k\,L_s^k\,a_k.
\eqno(42)$$
\medskip

The entities\enspace $b_n^+,\enspace n=\overline{0,\,N-1};\enspace r^+$
\enspace may be expressed in terms of thr right Bell polinomials if use the equality (12) and
take into account
\medskip

$$L_s^k\,a=L_s^k\,e\,a=(-1)^k\,B_k^+(s)\,a.$$
Hence links (41), (42) transform to:
\medskip

$$b_n^+=\sum_{k=n+1}^N\,B_{k-n-1}^+(s)\,a_k,\quad n=\overline{0,\,N-1};
\eqno(43)$$

$$r^+=\sum_{k=0}^N\,B_k^+(s)\,a_k.
\eqno(44)$$
\medskip

Formulas (30), (37), (43), (44) (or (41), (42)) give a solution of the left division problem of\,
$L$\, by
$L_s$\, .\, So , it is proved
\medskip

{\bf{Statement 11}}.\enspace {\it{if the representation}}\,
(31),\enspace {\it{is valid, then for the reminder}}\, $r^+$\, {\it{and the result}}\, $M^+$
\enspace {\it{the formulas}}\, (44)\, ({\it{or}}\, (42))\,
{\it{and}}\, (37),\, {\it{take place and for the coefficients of the operator}}\, $M^+$\,
{\it{there is a representation}}\, (43)\, ({\it{or}}\, (41)).
\medskip

The straight corollary of this sentence is the following
\medskip

{\bf{Statement 12}}.\enspace {\it{For the operator
 }}\, $L$\, {\it{to be left divisible by the operator} }\, $L_s$\, (without remainder)
{\it{it is necessary and enough}},\, {\it{that}}\, $s$\, {\it{be a solution of the
differential equation}}
\medskip

$$\sum_{k=0}^N\,B_k(s)^+\,a_k=0.
\eqno(45)$$

{\it{if this condition holds, then the operator}}\, $L$\,
{\it{factorizes as}}:
\medskip

$$L=L_s\,M^+,$$
{\it{where the value of the operator}}\, $M^+$\, {\it{is given by the expressions }}\,
(37), (43)\, ({\it{or}}\, (42)).
\medskip

The nonlinear equation (45) is called as \enspace {\it{generalized left Riccati equation}}.

The left Riccati equation is linearized obviously by the Statement 3. As a result we have
\medskip

{\bf{Statement \, 13}}.\enspace {\it{Let an invertible function}}\,
$\varphi$\, {\it{satisfy to the linear differential equation}}
\medskip

$$\sum_{n=0}^N\,(-1)^n\,B_n(s)^+\,a_n\,D^n\varphi=0.
\eqno(46)$$
{\it{Then the operator}}\, $L$,\, {\it{determined by the equality}}\, (31),\,
{\it{is left divisible by the operator}}\, $L_s$,\, {\it{where}}
\medskip

$$s=-\varphi^{-1}\,\varphi'.$$
\medskip

{\it{The function}}\, $s$\, {\it{is a solution to the left Riccati equation}}\, (45).
\medskip

{\bf{Remark\, 2}}.\enspace Following F. Calogero classification, the Riccati equations
\enspace  (38), (45) are\enspace
$C$--{\it{integrable dynamical systems}}.
\bigskip

{\bf{4.\enspace Darboux transformation.\enspace Generalized Burgers equations}}.
\medskip

We shall show that the problem under consideration of the operator division is connected 
directly with {\it{Darboux transformation}}.\enspace For the purpose we take a version in
which at the ring
$K$\, there exists one more differentiation\, $D_0$,\,which\enspace
{\it{commute}}\enspace with the operator D\, $D$,
\medskip

$$D_0\,D=D\,D_0.$$
\medskip

It may be a differentiation by a parameter \, $t$.
\medskip

Let us list auxiliary commutation relation.
\medskip

$$L_s\,r=r\,L_s+Dr+[r,\,s].
\eqno(47)$$
\medskip
Really,
\medskip

$$L_s\,r-r\,L_s=(D-s)\,r-r\,(D-s)=D\,r-sr-r\,D+rs=r\,D+Dr-sr-r\,D+rs=
Dr+[r,\,s].$$
\medskip

Taking into account the equalities (48) and (31), we calculate
\medskip

$$L_s\,(D_0-L)=L_s\,D_0-L_s\,L=D_0\,L_s+D_0s-L_s\,(M\,L_s+r)=
D_0\,L_s+D_0s-L_s\,M\,L_s-L_s\,r=$$
$$D_0\,L_s+D_0s-L_s\,M\,L_s-r\,L_s-Dr-[r,\,s]=
(D_0-L_s\,M-r)\,L_s+D_0s-Dr-[r,\,s].$$
\medskip

and arrive to the following relation:
$$L_s\,(D_0-L)=(D_0-\tilde L)\,L_s+D_0s-Dr-[r,\,s],
\eqno(48)$$
where
\medskip

$$\tilde L=L_s\,M+r.
\eqno(49)$$
\medskip

We results in the important conclusion:
\medskip

{\bf{Statement \, 14}}.\enspace {\it{If a function}}\, $s$\,
{\it{satisfy the equation}}
\enspace

$$D_0s=Dr+[r,\,s],
\eqno(50)$$
{\it{the operator}}\enspace $L_s$\enspace {\it{intertwine the operators}}
\enspace $D_0-L$\, 3\, $D_0-\tilde L$,
\medskip

$$L_s\,(D_0-L)=(D_0-\tilde L)\,L_s.
\eqno(54)$$
\medskip
Now we would obtain the explicit expression for\, $\tilde L$,\,in the terms of  (49), (31),
(32); namely
\medskip

$$\tilde L=L_s\,M+r=(D-s)\,\sum_{n=1}^N\,a_n\,H_{n-1}+
\sum_{n=0}^N\,a_n\,B_n(s)=$$
$$\sum_{n=1}^N\,D\,a_n\,H_{n-1}-\sum_{n=1}^N\,s\,a_n\,H_{n-1}+
\sum_{n=0}^N\,a_n\,B_n(s)=$$
$$\sum_{n=1}^N\,(Da_n\,H_{n-1}+a_n\,D\,H_{n-1}-s\,a_n\,H_{n-1}+
a_n\,B_n(s))+a_0\,B_0(s)=$$
$$\sum_{n=1}^N\,(Da_n\,H_{n-1}+a_n\,(H_n-B_n(s)+B_n(s))-s\,a_n\,H_{n-1})+
a_0\,e=$$
$$\sum_{n=1}^N\,(a_n'\,H_{n-1}+a_n\,H_n-s\,a_n\,H_{n-1})+a_0.$$
\medskip

Finally, the transfomed operator is
\medskip

$$\tilde L=\sum_{n=1}^N\,(a_n'\,H_{n-1}+a_n\,H_n-s\,a_n\,H_{n-1}))+a_0.
\eqno(52)$$
\medskip
Let us write the equation (50) in the explicit form by virtue of the formula (31).  
\medskip

$$Dr+[r,\,s]=D\,\sum_{n=0}^N\,a_n\,B_n(s)+
\left[\sum_{n=0}^N\,a_n\,B_n(s),\,s\right]=$$
$$\sum_{n=0}^N\,D\,a_n\,B_n(s)+\sum_{n=0}^N\,[a_n\,B_n(s),\,s]=$$
$$\sum_{n=0}^N\,(a_n'\,B_n(s)+a_n\,DB_n(s))+
\sum_{n=0}^N\,(a_n\,B_n(s)\,s-s\,a_n\,B_n(s))=$$
$$\sum_{n=0}^N\,(a_n'\,B_n(s)+a_n\,DB_n(s)+a_n\,B_n(s)\,s-
s\,a_n\,B_n(s))=$$
$$\sum_{n=0}^N\,(a_n'\,B_n(s)+a_n\,(DB_n(s)+B_n(s)\,s)-s\,a_n\,B_n(s))=
$$
$$\sum_{n=0}^N\,(a_n'\,B_n(s)+a_n\,B_{n+1}(s)-s\,a_n\,B_n(s)).$$
\medskip

It is established that for the intertwine relation (51) validity it is necessary and
enough that 
$s$\, should  be a solution of the equation
\medskip

$$D_0s=\sum_{n=0}^N\,(Da_n\,B_n(s)+a_n\,B_{n+1}(s)-s\,a_n\,B_n(s)).
\eqno(53)$$
\medskip

{\bf{Remark\, 3}}.\enspace The equation (53) is nonlinear, but linearizable, i.e. (by
Calogero classification) it is 
\, $C$--{\it{integrable.}}\, 
This equation (in different forms) was introduced at [12,16]. The form we suggest is most
compact nad convenient for a further investigations, e.g. in the franework of
bilineraization technique of [13].
\medskip

{\bf{Example}}.
\medskip

Let
\medskip

$$L=D^2.$$
Then
\medskip

$$\tilde L=H_2-s\,H_1=B_{2,\,0}(s)\,D^2+B_{2,\,1}(s)\,D+B_{2,\,2}(s)-
s\,(B_{1,\,0}(s)\,D+B_{1,\,1}(s))=$$
$$D^2+s\,D+(s^2+2Ds)-s\,(D+s)=D^2+2Ds.$$
\medskip

The equation (53) takes the form:
\medskip

$$D_0s=D^2s+2Ds\,s.$$
In the case of scalar functions it is known {\it{Burgers equation}}.\enspace By this
reason and due to the eqaution are integrable by the Cole-Hopf transformation, the equation
(53) is naturally named {\it{generalized Burgers equation}}.
\medskip

{\bf{Statement\, 15}}.\enspace {\it{Let an invertible function }}\,
$\varphi$,\, {\it{is a solution to linear
differential equation}}
\medskip

$$D_0\varphi=L\,\varphi.
\eqno(54)$$
{\it{Then the function}}\, $s$\, {\it{satisfy the generalized Burgers equation}}\, (56).
\medskip

{\bf{Proof}}.\enspace Let us note that
\medskip

$$(D\,r+[r,\,s])\,\varphi=(L_s)\,r\,\varphi+r\,s\,\varphi=
(L_s)\,r\,\varphi+r\,D\,\varphi=L_s\,(r\,\varphi)=L_s\,L\,\varphi.$$
Further, acting to the equation (53) by the operator\, $L_s$\, and
accounting for the relations (45) and (2), (16), we have
\medskip

$$0=L_s\,(D_0\varphi-L\,\varphi)=L_s\,D_0\varphi-L_s\,L\,\varphi=
(D_0\,L_s+D_0s)\,\varphi-L_s\,L\,\varphi=$$
$$D_0\,L_s\,\varphi+D_0s\,\varphi-(D\,r+[r,\,s])\,\varphi=
(D_0s-D\,r-[r,\,s])\,\varphi,$$
due to the existence of the inverse element for \, $\varphi$\, one obtain (50),
that is equivalent to  (52).\quad $\Box$
\medskip

The obvious corollary of the intertwine relation (51) and the statement16
is
\medskip

{\bf{Theorem (Matveev)}}.\enspace {\it{Let functions}}\enspace $\psi$\,
and\, $\varphi$\enspace {\it{are solutions of the equations}}
\medskip

$$D_0\,\psi=L\,\psi,\quad D_0\,\varphi=L\,\varphi,$$
{\it{for an invertible}}\, $\varphi$\, {\it{Then the function}}\,
\medskip

$$\tilde \psi=L_s\,\psi=D\,\psi - s\,\psi,\quad
s=D\,\varphi\,\varphi^{-1},
\eqno(55)$$
{\it{is a solution of the equation}}
\medskip

$$D_0\,\tilde \psi=\tilde L\,\tilde \psi.
\eqno(56)$$
\medskip

The last statement accomplishes the proof of Matveev theorem [15,16]. The equality (55) 
gives  a representation of the transformed operator in terms of the generalized Bell
polinomials. The explicit expression for the transformed coefficient is
$$
a_N[1] = a_N,
$$
$$
a_k[1] = a_k + \sum_{n=k}^N[a_nB_{n,n-k} + (a_n' - sa_n)B_{n-1,n-1-k}],\enspace k =
0,...,N-1.
$$

{\bf{5.\, Conclusion }}.
\medskip

It is shown that the division procedure for linear differential operators naturaly leads
to the solution of its factorisation problem thats links to intertwine relations and
Darboux transformations. The representation constructed here may, perhaps, open new
possibilities to build up and study new integrable systems.
One of us (S.leble) wish to thank TENA division of Brussels university (VUB) for
hospitality and F. Lambert for fruitful discussions.
Special gratitude authors bring to S.Tsarev for the priceless aid with copies of 
old-published papers and use of his unpublished paper [6].
The work is partially supported by Russian foundation for Basic Research, grant
N96-01-01789

\bigskip

{\bf{Appendix.\enspace Evaluation of generalized Bell polinomials.}}.
\medskip

Calculations by the relation (26) give the following:
\medskip

At\enspace $n=1$
\medskip

$$B_{n,\,1}(s)=\sum_{i=0}^0\,{n-i \choose -i}\,B_{n,\,i}(s)\,(D^{-i}s)=
{n \choose 0}\,B_{n,\,0}(s)\,(D^0s)=s;$$
\medskip

At\enspace $n=2$
\medskip

$$B_{n,\,2}(s)=\sum_{i=0}^1\,{n-i \choose 1-i}\,B_{n,\,i}(s)\,(D^{1-i}s)
={n \choose 1}\,B_{n,\,0}(s)\,(D^1s)+
{n-1 \choose 0}\,B_{n,\,1}(s)\,(D^0s)=$$
$$n\,Ds+s\,s=s^2+n\,Ds.$$
\medskip

At\enspace $n=3$
\medskip

$$B_{n,\,3}(s)=\sum_{i=0}^2\,{n-i \choose 2-i}\,B_{n,\,i}(s)\,(D^{2-i}s)
=$$
$${n \choose 2}\,B_{n,\,0}(s)\,D^2s+{n-1 \choose 1}\,B_{n,\,1}(s)\,D^1s+
{n-2 \choose 0}\,B_{n,\,2}(s)\,D^0s=$$
$${n \choose 2}\,D^2s+(n-1)\,s\,Ds+(s^2+n\,Ds)\,s=
s^3+n\,Ds\,s+(n-1)\,s\,Ds+{n \choose 2}\,D^2s.$$
\medskip

At\enspace $n=4$
\medskip

$$B_{n,\,4}(s)=\sum_{i=0}^3\,{n-i \choose 3-i}\,B_{n,\,i}(s)\,(D^{3-i}s)
=$$
$${n \choose 3}\,B_{n,\,0}(s)\,D^3s+{n-1 \choose 2}\,B_{n,\,1}(s)\,D^2s+
{n-2 \choose 1}\,B_{n,\,2}(s)\,D^1s+$$
$${n-3 \choose 0}\,B_{n,\,3}(s)\,D^0s=$$
$${n \choose 3}\,D^3s+{n-1 \choose 2}\,s\,D^2s+(n-2)\,(s^2+n\,Ds)\,Ds+$$
$$\left(s^3+n\,Ds\,s+(n-1)\,s\,Ds+{n \choose 2}\,D^2s\right)\,s=$$
$${n \choose 3}\,D^3s+{n-1 \choose 2}\,s\,D^2s+(n-2)\,s^2\,Ds+
n(n-2)\,(Ds)^2+s^4+n\,Ds\,s^2+$$
$$(n-1)\,s\,Ds\,s+{n \choose 2}\,D^2s\,s=$$
$$s^4+n\,Ds\,s^2+(n-1)\,s\,Ds\,s+(n-2)\,s^2\,Ds+{n \choose 2}\,D^2s\,s+
n(n-2)\,(Ds)^2+{n-1 \choose 2}\,s\,D^2s+$$
$${n \choose 3}\,D^3s.$$
\bigskip

1. Infeld L, Hull T.E. Rev. Mod. Phys. 1951 v. 23, p.21.

2.   K\"onigsberger, L.
, Allgemeine Untersuchungen aus der theorie
der Differentialgleichungen. Leipzig, Teubner, 1882, 246 p.

3. Umemura H.,
  Galois theory of algebraic and differential equations.
  {\em Nagoya Math. Journal}, 1996, v. 144, p. 1--58.

4. Andrianov A A, Borisov N V, Ioffe M V, Theor. Math. Phys., 1984,
v 61, p 121.

5. Grigor'ev, D.Yu.
   Complexity of factoring and calculating the
GCD of linear ordinary differential operators,
   {\em  J. Symbolic Computation 10\/} (1990), pp. 7--37.

6. Tsarev S.P. Applications of Factorization Methods to Integration of Nonlinear
 Ordinary and Partial Differential Equations, preprint 1999.

7. Landau E.
  Ein Satz \"uber die Zerlegung homogener linearer
Differentialausdr\"ucke in irreducible Factoren.
\newblock {\em J. f\"ur die reine und angewandte Math.}
(1901/1902), v. 124, p.115--120.

8. Loewy, A.
  {\"Uber} vollstandig reduzible lineare homogene
 Differentialgleichungen.
  {\em Math. Annalen 62\/} (1906), pp.89--117.
9. Ore, O.
  Linear equations in non-commutative fields.
  {\em Annals of Mathematics} (1931), v.32, p.463--477.

10. Tsarev, S. P.
  An algorithm for complete enumeration of all
factorizations of a linear ordinary differential operator.
 {\em Proceedings of ISSAC'96} (1996), ACM Press, p.226--231.

11. Matveev V B, Salle M A Darboux transformations and solitons
Springer-Verlag, Berlin, 1991.

12. Schimming R., Rida S.Z., Int. J. of Algebra and Computation, 1996. v.6, p.635-644. 

13. Gilson C, Lambert F., Nimmo J., Willox R., On the combinatorics of the Hirota D-operators
Proc. R. Soc. Lond A, (1996), v.452,p.223-234.

14. Lambert F. Loris I., Springael J. 1995 On a direct bilinearization method: Kaup
higher-order water wave equations as a modified nonlocal Boussinesq equation. J. Phys. A:
Math. Gen. {\bf 27} 5325.

15.  Matveev V.B. Lett. Math. Phys.(1979) {\bf 3} 213-216; 217-222; 503-512. Matveev V.B.
Salle M.A. Lett.Math.Phys. 1979, {\bf 3}, 425-429.

16. Matveev V.B., Hab. Diss. Theses, Leningrad,1980.

\vfill
\eject
\end{document}